\documentclass[grl]{AGUTeX}

\usepackage{graphicx}
\usepackage{amsmath}
\usepackage{color}

\setkeys{Gin}{draft=false}

\authorrunninghead{P\"AHTZ ET AL.}

\titlerunninghead{SATURATION OF AEOLIAN SAND TRANSPORT}

\authoraddr{Corresponding author: Thomas P\"ahtz, Institute of Physical Oceanography, Ocean College, Zhejiang University, 310058 Hangzhou, China. (0012136@zju.edu.cn)}

\begin{document}

%----------------------------------------------------------------------------------------
%	TITLE
%----------------------------------------------------------------------------------------

\title{Discrete Element Method simulations of the saturation of aeolian sand transport}

%----------------------------------------------------------------------------------------
%	AUTHORS AND AFFILIATIONS
%----------------------------------------------------------------------------------------

% Use \author{\altaffilmark{}} and \altaffiltext{}

% \altaffilmark will produce footnote; matching \altaffiltext will appear at bottom of page.

\authors{Thomas P\"ahtz,\altaffilmark{1,2}
Amir Omerad\v{z}i\'c,\altaffilmark{3}
Marcus V. Carneiro,\altaffilmark{3}
Nuno A. M. Ara\'ujo,\altaffilmark{4,5} and
Hans J. Herrmann\altaffilmark{3,6}}

\altaffiltext{1}{Institute of Physical Oceanography, Ocean College, Zhejiang University, 310058 Hangzhou, China}

\altaffiltext{2}{State Key Laboratory of Satellite Ocean Environment Dynamics, The Second Institute of Oceanography, 310012 Hangzhou, China}

\altaffiltext{3}{Institut f\"ur Baustoffe, ETH Z\"urich, 8093 Zurich, Switzerland}

\altaffiltext{4}{Departamento de F\'isica, Faculdade de Ci\^encias, Universidade de Lisboa, P-1749-016 Lisboa, Portugal}

\altaffiltext{5}{Centro de F\'isica Te\'orica e Computacional, Universidade de Lisboa, Avenida Professor Gama Pinto 2, P-1649-003 Lisboa, Portugal}

\altaffiltext{6}{Departamento de F\'isica, Universidade Federal do Cear\'a, 60451-970 Fortaleza, Cear\'a, Brazil}

%----------------------------------------------------------------------------------------
%	ABSTRACT
%----------------------------------------------------------------------------------------

% Do NOT include any \begin...\end commands within the body of the abstract.

\begin{abstract}

The saturation length of aeolian sand transport ($L_s$), characterizing the distance needed by wind-blown sand to adapt to changes in the wind shear, is essential for accurate modeling of the morphodynamics of Earth's sandy landscapes and for explaining the formation and shape of sand dunes. In the last decade, it has become a widely-accepted hypothesis that $L_s$ is proportional to the characteristic distance needed by transported particles to reach the wind speed (the ``drag length''). Here we challenge this hypothesis. From extensive numerical Discrete Element Method simulations, we find that, for medium and strong winds, $L_s\propto V_s^2/g$, where $V_s$ is the saturated value of the average speed of sand particles traveling above the surface and $g$ the gravitational constant. We show that this proportionality is consistent with a recent analytical model, in which the drag length is just one of four similarly important length scales relevant for sand transport saturation.

\end{abstract}

\begin{article}

\section{Introduction}
Aeolian transport of sand occurs when a sufficiently strong wind blows over a sand bed \citep{Bagnold41,Shao08,Duranetal11,Koketal12}. The two dominant transport modes are saltation, referring to particles hopping along the sand surface in characteristic trajectories \citep{Bagnold41}, and creep, referring to particles rolling and sliding along the sand surface \citep{Bagnold37}. Wind-blown, initially flat sand beds may evolve into bedforms, such as ripples and dunes, due to different kinds of instabilities \citep{ClaudinAndreotti06,Andreottietal10,Fourriereetal10,Partelietal11,Charruetal13,Duranetal14b}.

For instance, dunes are thought to form due to an aerodynamic instability, namely a slight phase difference between topography and wind shear maxima on a periodically perturbed sand bed \citep{JacksonHunt75,Huntetal88,Kroyetal02a,Kroyetal02b}. If this phase difference is larger than the phase difference between sand transport and wind shear maxima, which is characterized by the saturation length ($L_s$) \citep{Sauermannetal01,ParteliHerrmann07b,Andreottietal10,Paehtzetal13,Paehtzetal14}, the perturbations grow. This is one of the reasons why $L_s$ plays an important role in aeolian dune formation. Indeed, $L_s$ controls the length of the smallest (``elementary'') dunes evolving from a flat sand bed and the minimal size of crescent-shaped barchans \citep{Partelietal07,ClaudinAndreotti06,Fourriereetal10}. In contrast, the steady state dune dimensions are controlled by the aerodynamic roughness ($z_o^*$) \citep{Pelletier09}. $L_s$ is also a key parameter in morphodynamic models of Earth's sandy landscapes, such as aeolian dune models \citep{Kroyetal02a,Kroyetal02b,SchwaemmleHerrmann03,ParteliHerrmann07a,Narteauetal09,Partelietal09,Partelietal14}.

It has been a challenging task to predict $L_s$ as a function of wind and particle parameters, such as the wind shear velocity ($u_\ast$), the mean particle diameter ($d$), the particle ($\rho_p$) and fluid density ($\rho_f$), and the kinematic air viscosity ($\nu$). In fact, the main difficulty has been to understand which of the involved relaxation mechanism is the slowest and thus the most important one. \citet{Sauermannetal01} derived an expression for $L_s$ based on the assumption that $L_s$ corresponds to the length needed to eject particles from the sand bed. Based on measurements of the size of both subaqueous and aeolian barchan dunes, \citet{Hersenetal02} proposed that $L_s$ is proportional to the drag length ($L_d=(\rho_p/\rho_f)d$), which characterizes the distance transported particles need to reach the flow speed. Estimations of the wavelength of elementary dunes by \citet{ClaudinAndreotti06} and measurements by \citet{Andreottietal10} later supported this proposition. Indeed, these measurements confirmed that $L_s$ is approximately proportional to $d$ and essentially independent of $u_\ast$, as predicted by $L_s\propto L_d$. However, there is considerable room for alternative interpretations of these measurements. The analytical model for the saturation length of both subaqueous and aeolian particle transport recently proposed by \citet{Paehtzetal13,Paehtzetal14} is also consistent with these measurements (without fitting), even though the model predicts that $L_s$ varies with $u_\ast$. In this model, the four potentially most important relaxation mechanisms are all accounted for (ejection of bed particles and particle deceleration in particle-bed collisions, fluid drag acceleration of particles, relaxation of the fluid speed), and it turns out that neglecting any of them entirely changes the model predictions \citep{Paehtzetal14}. This shows that the identity of the most important relaxation mechanisms remains an open problem.

Here we use Discrete Element Method (DEM) simulations for the particle phase to investigate aeolian sand transport saturation. This modeling technique considers interparticle interactions above and also with the sand bed and is thus more realistic than older modeling techniques \citep[e.g.,][]{Almeidaetal07,Almeidaetal08,KokRenno09}, which usually consider the sand bed as a flat, rough wall. However, it is also computationally more costly, which is the main reason why this technique had not been used for modeling particle-laden flows until a few years ago \citep{Carneiroetal11,Duranetal12,Carneiroetal13,Duranetal14a,Duranetal14b,Schmeeckle14,Paehtzetal15}. From our simulations, we find that the total mass of particles transported above the sand bed ($M$) relaxes significantly slower towards its saturated value ($M_s$) than the average particle velocity above the sand bed ($V=Q/M$, where $Q$ is the sand transport rate above the sand bed) towards its saturated value ($V_s$), indicating that the drag length is not the dominant saturation length scale. Moreover, we find that $L_s\propto V_s^2/g$, where $g$ is the gravitational constant, when $u_\ast>4u_t$, where $u_t$ is the dynamic threshold of sand transport (i.e., the extrapolated value of $u_\ast$ at which the saturated sand transport rate ($Q_s$) vanishes). This finding is consistent with the analytical model by \citet{Paehtzetal13,Paehtzetal14}, supporting the hypothesis that the aforementioned four potentially most important relaxation mechanisms are all similarly relevant.

This paper is organized as follows. First, we present the modeling technique we used to simulate aeolian particle transport in Section~\ref{Model}. Afterwards we show our numerical results in Section~\ref{Results}, which are then discussed and compared with the analytical model for the saturation length recently proposed by \citet{Paehtzetal13,Paehtzetal14} in Section~\ref{Discussion}. Finally, we draw conclusions in Section~\ref{Conclusion}.

\section{Modeling technique} \label{Model}
In this section, we briefly describe the three-dimensional numerical model which we used to model sand transport. A more detailed description can be found in \citet{Carneiroetal13}, particularly its supplementary material.

For the computation of the mean horizontal wind velocity ($u$), the model uses the mixing-length approximation of the Reynolds-averaged Navier-Stokes equations (neglecting viscous wind shear), where the mixing length is given by $\kappa(z-h)$ with $\kappa=0.4$ being the von Karm\'an constant and $z-h$ characterizing the vertical distance from the top of the sand bed ($z=h$). This reads \citep{Carneiroetal13}
\begin{eqnarray}
 \frac{\mathrm{d}u}{\mathrm{d}z}=\frac{u_\ast}{\kappa(z-h)}\sqrt{1-\frac{1}{\rho_fu_\ast^2}\int\limits_z^\infty f\mathrm{d}z}. \label{windprofile}
\end{eqnarray}
The term containing $f$, the horizontal drag force per unit volume applied by the wind on the particles (the drag law by \citet{Cheng97} is used), takes into account that the wind speed is reduced due to continuous transfer of momentum from wind to particles. The integration of Eq.~(\ref{windprofile}) starts at height $z=h+z_o$, where $z_o=d/30$ is the aerodynamic roughness of the sand bed as it would be in the absence of sand transport, corresponding to aerodynamically rough flow \citep{Bagnold41}. However, during sand transport, the reduced wind speed at the top of the saltation layer corresponds to an increased roughness value ($z_o^*$).

It is important to note that Eq.~(\ref{windprofile}) is applied to calculate the wind velocity profile at every time step because we assume that the flow adapts to local drag decelerations of the wind speed within the integration time ($\Delta t=0.005$ s). This standard assumption led to several previous numerical results in agreement with experiments \citep{Carneiroetal11,Duranetal12,Carneiroetal13,Duranetal14a,Paehtzetal15}. Nevertheless, we argue why it is reasonable to make such assumption in the following.

There are actually two time scales involved. First, the time needed to transmit the drag force between fluid and particles. As the drag force is transmitted via collisions between air molecules (which are extremely small) and the sand grains, this time scale is much smaller than the integration time. The second time scale is related to the propagation of this perturbation to the entire system. As perturbations of a fluid typically travel at the speed of sound (in air, $c\approx340$ m/s) and the maximum distance they need to travel to reach all relevant locations of the simulated saltation layer is of the order of $100d=0.02$ m (the height of the saltation layer), the maximal time needed for this perturbation to influence all relevant locations of the simulated saltation layer is around $0.0006$ s, which is around a factor $10$ smaller than $\Delta t$ and around a factor $10^3$ smaller than the saturation time. Even if the necessary time for the flow to accommodate to a perturbation is larger than the necessary time a perturbation needs to travel to reach all relevant locations, it is hard to imagine that this time comes any close to the saturation time. Moreover, since particle and flow velocity are of the same order of magnitude, also the length needed for the flow to adapt to the perturbation should be much smaller than the saturation length.

Trajectories and velocities of particles are obtained from solving Newton's equations of motion through the velocity-St\"ormer-Verlet scheme \citep{Griebeletal07}, considering gravity and wind drag as the external forces acting on the particles. Interparticle contacts are modeled through a dissipative spring dashpot potential (coefficient of restitution, $e=0.65$), while frictional contacts are neglected (no particle rotation). The system dimensions are length $\times$ height $\times$ width $=50d\times400d\times7.5d$, and $1410$ particles with normally distributed diameters ($d_p=(1\pm0.1)d$) are simulated. Most of these particle constitute a bed of around twelve particle layers. This is sufficiently thick to suppress the reflection of shock waves from the dissipative ($e=0.5$) bottom wall \citep{Rioualetal00,Rioualetal03}. The simulation top is open and the side boundaries periodic. In fact, particles never reach the top of the system.

\section{Results} \label{Results}
Using the model described in Section~\ref{Model}, we carried out simulations for typical sand transport conditions on Earth ($g=9.81$ m/s$^2$, $d=200$ $\mu$m, $\rho_p=2650$ kg/m$^3$, $\rho_f=1.174$ kg/m$^3$, $\nu=1.59\times10^{-5}$ m$^2$/s). For these conditions, we varied $u_\ast$ between two and nearly ten times the threshold shear velocity ($u_t=0.195$ m/s). For each $u_\ast$, $15$ runs were performed, starting from different initial condition. Indeed, while the sand bed in all samples was exactly the same at the start of each simulation ($t=0$), ten particles with velocity $v_x=v_z=1$ m/s were randomly placed sufficiently high above the surface ($z>h+20d$). Each of these samples evolved in time towards the saturated state. The supplementary online material contains a movie showing the time evolution of a given sample ($u_\ast=0.8$ m/s) between $t=0$ and $t=0.35$ s.

\begin{figure}
 \begin{center}
  \includegraphics[width=1.0\columnwidth]{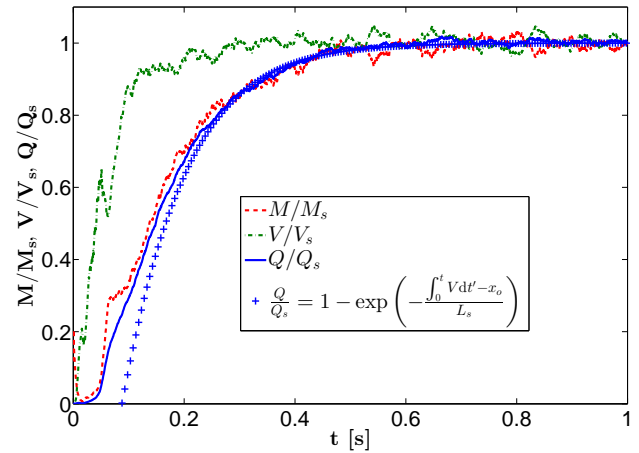}
 \end{center}
 \caption{Time evolution of $M/M_s$ (red, dashed line), $V/V_s$ (green, dash-dotted line), and $Q/Q_s$ (blue, solid line) obtained from simulations of typical sand transport conditions on Earth ($g=9.81$ m/s$^2$, $d=200$ $\mu$m, $\rho_p=2650$ kg/m$^3$, $\rho_f=1.174$ kg/m$^3$, $\nu=1.59\times10^{-5}$ m$^2$/s) with $u_\ast=1.2$ m/s. The blue crosses correspond to the best fit of Eq.~(\ref{Saturationprofile}) to the data $[Q/Q_s](t)$ near saturation ($|1-Q/Q_s|<0.2$). The analogous plots for different values of $u_\ast$ look similar.}
 \label{Profiles}
\end{figure}

From averaging the particle locations and velocities over the $15$ samples corresponding to each $u_\ast$ and further over the horizontal ($x$) and lateral ($y$) direction, we obtain the vertical profiles of the local mass density ($\rho$) and the mass-weighted average particle velocity $\langle\mathbf{v}\rangle$ the particle velocity at each time. The time evolution of $M=\int_h^\infty\rho\mathrm{d}z$ (red, dashed line), $Q=\int_h^\infty\rho\langle v_x\rangle\mathrm{d}z$ (blue, solid line), and $V=Q/M$ (green, dash-dotted line) further obtained from these profiles relative to their saturated values is plotted in Fig.~\ref{Profiles} for $u_\ast=1.2$ m/s (for different $u_\ast$, it looks similar).
It can be seen that the time transient behavior of $Q$ is similar to that determined in older studies \citep{Spiesetal00,MaZheng11}. Furthermore, one immediately recognizes that $M$ relaxes significantly slower towards $M_s$ than $V$ towards $V_s$ (this is also true for our simulations with other values of $u_\ast$). This means that our simulations do not confirm the hypothesis that drag is the dominant mechanism controlling sand transport saturation, which would instead require that $M$ always relaxes much faster towards $M_s$ than $V$ towards $V_s$ \citep{Paehtzetal14}. Moreover, this observation can be used to extract the saturation length ($L_s$) from the time evolution of $Q$, as we explain in the following.

Mathematically, $L_s$ has its origin in the mass conservation equation, which in its local form reads
\begin{eqnarray}
 \frac{\partial\rho}{\partial t}+\frac{\partial\rho\langle v_x\rangle}{\partial x}+\frac{\partial\rho\langle v_y\rangle}{\partial y}+\frac{\partial\rho\langle v_z\rangle}{\partial z}=0. \label{massconserv}
\end{eqnarray}
Indeed, $L_s$ is typically defined through a first-order Taylor expansion of the rate of relaxation ($\Gamma(Q)$) of $Q$ around $Q_s$ \citep{Andreottietal10,Paehtzetal13,Paehtzetal14},
\begin{eqnarray}
 \frac{\mathrm{d}Q}{\mathrm{d}x}=\Gamma(Q)\simeq\frac{Q_s-Q}{L_s}, \label{relaxation}
\end{eqnarray}
which, using $\Gamma=[\rho\langle v_z\rangle](h)$, is the height integration ($\int_h^\infty\cdot\mathrm{d}z$) of Eq.~(\ref{massconserv}) for steady and laterally homogeneous conditions ($\partial/\partial t=\partial/\partial y=0$). Since $\Gamma(Q_s)=0$, $L_s$ corresponds to the negative inverse Taylor coefficient ($-[\Gamma^\prime(Q_s)]^{-1}$). By definition Eq.~(\ref{relaxation}), describing the spatial relaxation of $Q$ towards $Q_s$, is only applicable near saturation ($|1-Q/Q_s|\ll1$).

Since our simulations correspond to spatially and laterally homogeneous conditions ($\partial/\partial x=\partial/\partial y=0$), height integration of Eq.~(\ref{massconserv}) yields
\begin{eqnarray}
 \frac{\mathrm{d}Q}{V\mathrm{d}t}\simeq\frac{\mathrm{d}M}{\mathrm{d}t}=\Gamma(Q)\simeq\frac{Q_s-Q}{L_s}, \label{relaxationsim}
\end{eqnarray}
where the approximation on the left hand side uses that $M$ relaxes significantly slower towards $M_s$ than $V$ towards $V_s$. From comparison between Eqs.~(\ref{relaxation}) and (\ref{relaxationsim}), it is apparent that the saturation in $t$ in our simulations is equivalent to a saturation in $x$ if $x=\int_0^tV\mathrm{d}t'$ (i.e., $\mathrm{d}/\mathrm{d}x=V\mathrm{d}/\mathrm{d}t$) is used to relate them. In fact, fitting (nonlinear least squares method) $Q_s$, $x_o$, and $L_s$ to best agreement with the analytic solution
\begin{eqnarray}
 \frac{Q}{Q_s}=1-\exp\left(-\frac{\int_0^tV\mathrm{d}t'-x_o}{L_s}\right) \label{Saturationprofile}
\end{eqnarray}
of Eq.~(\ref{relaxationsim}) (since $Q_s$ does not depend on $t$ for constant wind shear) allows us to determine $L_s$ from our simulations (see blue crosses in Fig.~\ref{Profiles}).

\begin{figure}
 \begin{center}
  \includegraphics[width=1.0\columnwidth]{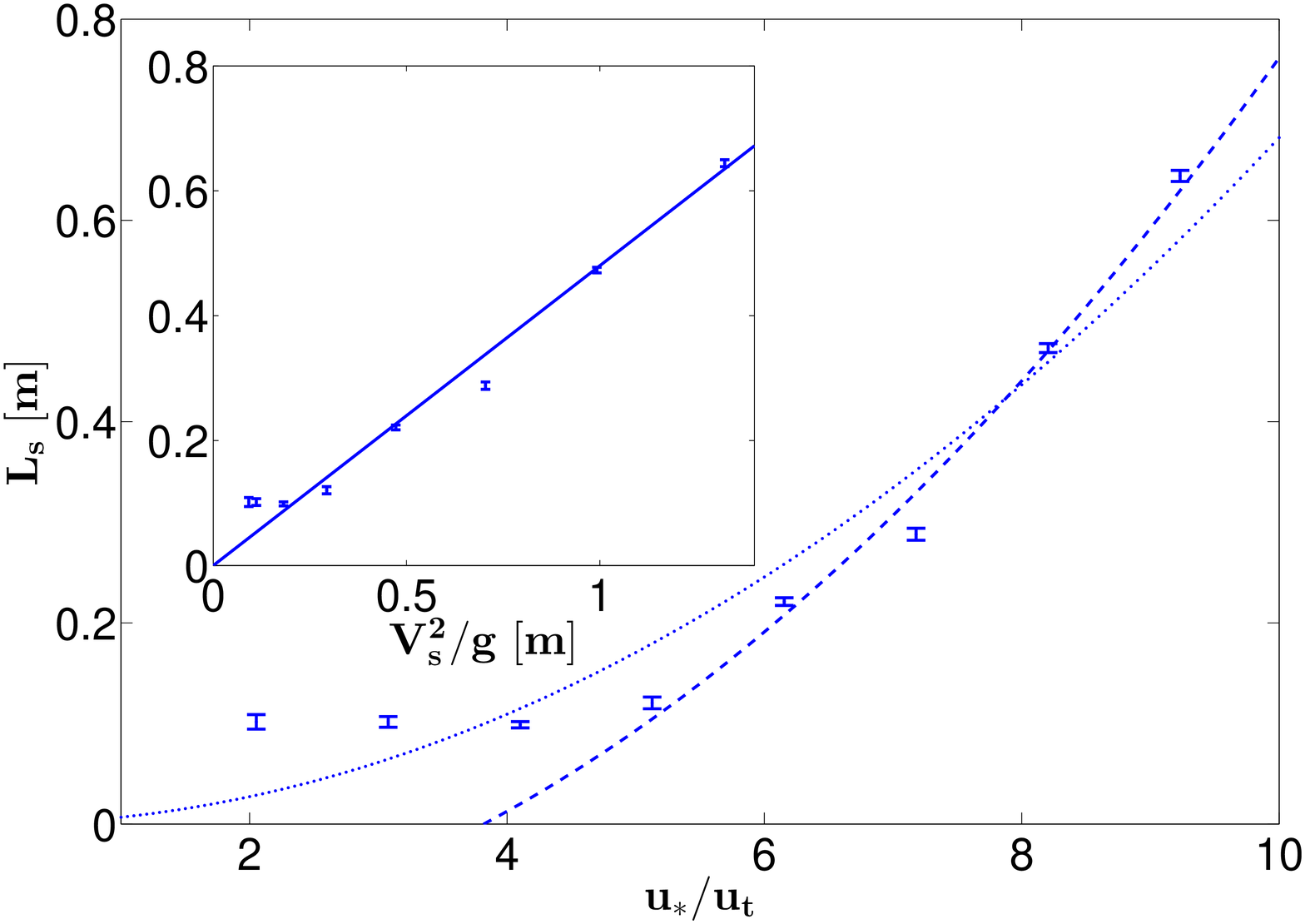}
 \end{center}
 \caption{The saturation length ($L_s$) as a function of $u_\ast/u_t$ (symbols) obtained from simulations of typical sand transport conditions on Earth ($g=9.81$ m/s$^2$, $d=200$ $\mu$m, $\rho_p=2650$ kg/m$^3$, $\rho_f=1.174$ kg/m$^3$, $\nu=1.59\times10^{-5}$ m$^2$/s), whereby the dashed line corresponds to $L_s=2.30u_\ast^2/g-0.13$ m and the dotted line to $L_s=1.76u_\ast^2/g$. The inset shows $L_s$ as a function of $V_s^2/g$ for the same conditions, whereby the solid line corresponds to $L_s=0.48V_s^2/g$. The error bars correspond to the $95\%$-confidence intervals obtained from the best fits to Eq.~(\ref{Saturationprofile}).}
 \label{Lsfig}
\end{figure}

Fig.~\ref{Lsfig} shows $L_s$ as a function of $u_\ast/u_t$ obtained from our simulations, whereby the error bars correspond to the $95\%$-confidence intervals obtained from the best fits to Eq.~(\ref{Saturationprofile}).
It can be seen that $L_s$ remains nearly constant between $2u_t$ and $4u_t$, qualitatively consistent with measurements \citep{Andreottietal10}. However, $L_s$ increases with $u_\ast$ when $u_\ast>4u_t$. This increase within the error bars follows the scaling relation
\begin{eqnarray}
 L_s=\alpha V_s^2/g, \label{Lsscaling}
\end{eqnarray}
as shown in the inset of Fig.~\ref{Lsfig}, where $\alpha=0.48$. Interestingly, while the scaling $L_s\propto V_s^2/g$ describes the data with $u_\ast>4u_t$ very well, the scaling $L_s\propto u_\ast^2/g$ does not (see the dotted line in Fig.~\ref{Lsfig}, corresponding to the best fit of $L_s\propto u_\ast^2/g$ to data with $u_\ast/u_t>4$), indicating that $V_s$ and not $u_\ast$ is the relevant parameter controlling $L_s$. Only when one allows a small offset, a good fit can be obtained. This is shown by the dashed line in Fig.~\ref{Lsfig}, which corresponds to $L_s+L_{so}\propto u_\ast^2/g$, where the offset ($L_{so}=0.13$ m) is expected to have a complex dependency on particle and wind parameters (except $u_\ast$).

\section{Discussion} \label{Discussion}
In this section, we first briefly describe the analytical model for the saturation length proposed by \citet{Paehtzetal13,Paehtzetal14} in Section~\ref{Anamodel}. We then compare the model predictions with our numerical results shown in Fig.~\ref{Lsfig} and with the scaling in Eq.~(\ref{Lsscaling}).

\subsection{Analytical model by \citet{Paehtzetal13,Paehtzetal14}} \label{Anamodel}
The analytical model for the saturation length by \citet{Paehtzetal13,Paehtzetal14} takes into account that $M$ and $V$ relax towards $M_s$ and $V_s$, respectively, due to different relaxation mechanisms. Changes in $M$ are controlled by the ejection of bed particles in particle-bed collisions \citep{Koketal12}, while changes in $V$ are driven by the acceleration of transported particles due to fluid drag and their deceleration in particle-bed collisions. On the one hand, the changes of $V$ were explicitly modeled within the momentum balance: the fluid drag acceleration through the fluid drag law by \citet{Julien95} for natural sand and the deceleration in particle-bed collisions by means of a Coulomb friction law, assuming a proportionality between average horizontal and vertical forces acting on transported particles. This assumption is well established in the literature as it leads to the experimentally \citep{Creysselsetal09} and numerically \citep{KokRenno09,Paehtzetal12} confirmed relation
\begin{eqnarray}
 M_s=\frac{\rho_f}{\mu g}(u_\ast^2-u_t^2), \label{Ms}
\end{eqnarray}
where $\mu\approx1$ (from experimental data \citep{Creysselsetal09,Paehtzetal12}) is the associated Coulomb friction coefficient. On the other hand, changes in $M$ were not explicitly modeled, but implicitly accounted for in the parameter
\begin{eqnarray}
 c_M=\frac{V_s}{M_s}\frac{\mathrm{d}M}{\mathrm{d}V}(V_s),
\end{eqnarray}
which describes the relative change of $M$ with $V$ near the saturated regime \citep{Paehtzetal14}. Moreover, taking into account that not only $M$ and $V$, but also the average wind speed ($U$) relaxes towards its saturated value, led to the introduction of another parameter $c_U$, describing the relative change of $U$ with the bed fluid shear velocity (i.e., the value of the fluid shear velocity at the bed, which is smaller than $u_\ast$ due to momentum transfer from fluid to particles). The parameters $c_M$ and $c_U$ were by far the most uncertain model parameters as they were the only ones not determined by measurements, but instead by theoretical arguments (which led to $c_M\approx c_U\approx 1$ for aeolian sand transport) \citep{Paehtzetal13,Paehtzetal14}. However, in the limit of large fluid shear velocities ($u_\ast/u_t\gg1$), the model becomes independent of $c_U$. Indeed, in this limit, the final model equation for $L_s$ reads \citep{Paehtzetal14},
\begin{eqnarray}
 L_s=\frac{(2+c_M)c_v}{c_M\mu}\frac{V_s^2}{g}, \label{ModelLs}
\end{eqnarray}
where $c_v\approx1.3$ (from experimental data \citep{RasmussenSorensen08,Creysselsetal09,Paehtzetal13,Paehtzetal14}) is the saturated value of $\int_h^\infty\rho\langle v_x^2\rangle\mathrm{d}z/(MV^2)$.

\subsection{Comparison between analytical and numerical model predictions} \label{Compmodel}
It can be seen that Eq.~(\ref{ModelLs}) predicts $L_s\propto V_s^2/g$, which resembles the scaling in Eq.~(\ref{Lsscaling}) obtained from our simulations. Since both equations are only valid for sufficiently large values $u_\ast/u_t$, this resemblance supports the analysis by \citet{Paehtzetal13,Paehtzetal14}. In this analysis, the four potentially most important relaxation mechanisms are all accounted for (ejection of bed particles and particle deceleration in particle-bed collisions, fluid drag acceleration of particles, relaxation of the fluid speed), as described in Section~\ref{Anamodel}. Since neglecting any of them entirely changes the model predictions \citep{Paehtzetal14}, the resemblance between the analytical and numerical model predictions suggests that these four relaxation mechanisms are all similarly relevant. This is just another indication that $L_d$ is {\em{not}} the dominant length scale controlling sand transport saturation.

However, one must also note that there are differences between these models. First, the qualitative model predictions for small $u_\ast/u_t$ slightly differ from each other. While the numerical model predicts that $L_s$ remains nearly constant between $2u_t$ and $4u_t$, the analytical model predicts a slight increase with $u_\ast$ \citep{Paehtzetal13,Paehtzetal14}. This might be a result of the aforementioned uncertainty of the parameter $c_U$. Second, while the analytical model predictions are consistent with the measurements by \citet{Andreottietal10}, the numerical model predictions shown in Fig.~\ref{Lsfig} are not. This can be entirely linked to differences in the model parameters $\mu$ and $c_M$, as we explain in the following. First, from fitting Eq.~(\ref{Ms}) to our numerical data (it fits very well, not shown), we obtain $\mu\approx2$, in contrast to $\mu\approx1$, which \citet{Paehtzetal13,Paehtzetal14} obtained from experimental data. Second, since $M$ relaxes much slower towards $M_s$ than $V$ towards $V_s$ in the simulations (see Fig.~\ref{Profiles}), $c_M$ becomes very large, while $c_M\approx1$ was estimated by \citet{Paehtzetal13,Paehtzetal14} from theoretical arguments. In fact, using $\mu\approx2$ and $c_M\rightarrow\infty$ in Eq.~(\ref{ModelLs}) yields the prefactor $(2+c_M)c_v/(c_M\mu)\rightarrow\approx0.65$ close to the prefactor $\alpha=0.48$ in Eq.~(\ref{Lsscaling}). This means the numerical model and the analytical model seem quantitatively consistent with each other since the differences in the parameters $\mu$ and $c_M$ are likely the results of simplifications in the numerical model. For instance, the difference in the value of $\mu$ between simulations and measurements can be linked to the interparticle contact model, which is known to have considerable influence on the frictional behavior of solids \citep{Campbell06}. Indeed, the model neglects particle rotation and uses rather soft particles (stiffness $k=1$kg/s$^2$), which allows particle overlaps of about $20\%$, while in reality the stiffness is several orders of magnitude larger. Also, the coefficient of restitution used in our simulations ($e=0.65$) might have been too small. Comparable studies usually use larger values \citep[e.g.,][$e=0.9$]{Duranetal12} and obtain a Coulomb friction coefficients near unity.

\section{Conclusion} \label{Conclusion}
We simulated aeolian sand transport using DEM simulations. From these simulations, we obtained the saturation curves in Fig.~\ref{Profiles} of the total mass of particles transported above the sand bed ($M$), their average velocity ($V$), and the associated sand flux ($Q=MV$). These numerical data indicate that $M$ saturates much slower than $V$, challenging the widely-accepted hypothesis that the drag length ($L_d=(\rho_p/\rho_f)d$) is the dominant length scale controlling aeolian sand transport saturation, which would require the opposite, namely that $M$ saturates much faster than $V$. Since $L_d$ does not change with $u_\ast$, the same hypothesis is further challenged by the numerical data in Fig.~\ref{Lsfig} showing that the saturation length ($L_s$) significantly increases with the wind shear velocity ($u_\ast$) for medium and strong winds ($u_\ast>4u_t$). Moreover, this increase follows the scaling relation $L_s\propto V_s^2/g$ (see inset of Fig.~\ref{Lsfig}), qualitatively consistent with the limit $u_\ast/u_t\gg1$ of the recently proposed analytical model by \citet{Paehtzetal13,Paehtzetal14}. In Section~\ref{Discussion}, we showed that this analytical model also predicts the proportionality factor in $L_s\propto V_s^2/g$ to be about $0.65$, which is close to the numerically obtained value $\alpha=0.48$. This adds another piece of doubt on a dominating role of $L_d$ because the model accounts for the four potentially most important relaxation mechanisms (ejection of bed particles and particle deceleration in particle-bed collisions, fluid drag acceleration of particles, relaxation of the fluid speed), and neglecting any of them entirely changes the model predictions \citep{Paehtzetal14}.

The scaling relation $L_s\propto V_s^2/g$, found for medium and strong winds, might itself become an important step towards modeling of dune and dune field evolutions in sand storms. For this purpose, one would need a model predicting $V_s$. In fact, there are a considerable number of analytical models predicting $V_s$ as function of $u_\ast$ \citep[e.g.,][]{Bagnold41,Kawamura51,Owen64,Bagnold73,Kind76,LettauLettau78,UngarHaff87,Sorensen91,Sorensen04,Duranetal11,Paehtzetal12,Laemmeletal12}, some of them might be applicable to sand storm conditions.

Finally, it is worth to note that aeolian dunes are often superimposed by ripples. Compared to the flat sand bed condition present in our simulation, the presence of such ripples leads to a strong increase of the aerodynamic roughness ($z_o^*$), which corresponds to a smaller wind and thus saturated particle velocity ($V_s$) in the saltation layer. The relation $L_s\propto V_s^2/g$, found for medium and strong winds, would thus imply that the formation of superimposed ripples on the surface of aeolian dunes is associated with a simultaneous decrease of $L_s$.

\begin{acknowledgments}
The data displayed in Figs.~\ref{Profiles} and \ref{Lsfig} are available from the authors. This work was partially supported by the National Natural Science Foundation of China (Grant No. 41350110226), the Brazilian Council for Scientific and Technological Development CNPq, ETH (Grant No. ETH-10 09-2), the European Research Council (Grant No. FP7-319968), and the Portuguese Foundation for Science and Technology (FCT) under Contracts nos. EXCL/FIS-NAN/0083/2012, PEst-OE/FIS/UI0618/2014, and IF/00255/2013. 
\end{acknowledgments}

%\bibliographystyle{agufull08}
%\bibliography{model}

\end{article}
%\newpage

\end{document}